\documentclass[fleqn,12pt,twoside]{article}
\usepackage{espcrc1,amssymb}

\usepackage{graphicx}
\usepackage[figuresright]{rotating}

\newcommand{\xbj}{x_{\mbox {\scriptsize Bj}}}
\newcommand{\AmS}{{\protect\the\textfont2
  A\kern-.1667em\lower.5ex\hbox{M}\kern-.125emS}}

\title{Deeply Virtual Compton Scattering on the Deuteron
\thanks{Work supported by the EC--IHP Network ESOP, 
Contract HPRN-CT-2000-00130.}}

\author{F. Cano\address{DAPNIA/SPhN, CEA--Saclay, F91191
Gif-sur-Yvette Cedex, France} and B. Pire\address{CPhT, 
\'Ecole Polytechnique, UM C7644 of CNRS ,F-91128 Palaiseau, France}}

\begin{document}

% typeset front matter
\maketitle

\begin{abstract}
We study deeply virtual Compton scattering (DVCS) on a deuteron
target. We model the Generalized Quark Distributions in the deuteron by
using the impulse approximation for the lowest Fock-space
state. Numerical predictions are given for the unpolarized cross
sections for the kinematical regimes relevant for JLab and HERMES at HERA. Differential cross sections show the same pattern as for the proton
case and at low values of $-t$ they are of comparable size.
 \end{abstract}

\section{INTRODUCTION}

	In recent years it has become clear that hard exclusive
processes, such as DVCS and deeply exclusive meson production (DEMP),
play a unique role in offering a rather complete picture of the
hadronic structure (for a recent review see \cite{GOEKE01} and
references therein). The
information which can be accessed through these experiments is encoded
by the Generalized Parton Distributions (GPDs), whose physical
interpretation has been elucidated by some authors
\cite{RADYUSHKIN97}.  Recent measurements of the azimuthal dependence of
the beam spin asymmetry in DVCS \cite{HERMES01,CLAS01} have provided
experimental evidence to support the validity of the formalism of GPDs
and the underlying QCD factorization theorems.

	The theoretical arguments used in deriving factorization
theorems in QCD for the nucleon can be applied for the deuteron as well, and
therefore one can develop the formalism of GPDs for the deuteron. From
the theoretical viewpoint, it is the simplest and best known nuclear
system and represents the most appropriate starting point to
investigate hard exclusive processes off nuclei. On the other hand,
these processes could offer a new source of information about the
partonic degrees of freedom in nuclei, complementary to the existing
one from deep inelastic scattering. Experimentally, deuteron targets
are quite common and as a matter of fact DVCS experiments are 
planned or being carried out at facilities like JLab and HERA. 

A parameterization of the non-perturbative matrix
elements which determine the amplitudes in DVCS and DEMP on a spin-one
target were given in terms of nine GPDs for the quark sector
  \cite{BERGER01}:   

\begin{eqnarray}
  \label{vaten}
V_{\lambda'\lambda} &=&
  \int \frac{d \kappa}{2 \pi}\,
  e^{i x \kappa 2 \bar{P}.n}
  \langle P', \lambda' |\,
  \bar{\psi}(-\kappa  n)\, \gamma.n\, \psi(\kappa n)
  \,| P, \lambda \rangle
%\nonumber \\
= \sum_{i=1,5}
  \epsilon'^{\ast \beta}  V^{(i)}_{\beta \alpha}\,
  \epsilon^{\alpha}\, H_{i}(x,\xi,t) ,
\\
A_{\lambda'\lambda} &=&
  \int \frac{d \kappa}{2 \pi}\,
  e^{i x \kappa 2 \bar{P}.n}
  \langle P', \lambda' |\,
  \bar{\psi}(-\kappa n)\, \gamma.n \gamma_5\, \psi(\kappa n)
  \,| P, \lambda \rangle
 =  \sum_{i=1,4}
  \epsilon'^{\ast \beta}  A^{(i)}_{\beta \alpha}\,
  \epsilon^{\alpha}\, \tilde{H}_{i}(x,\xi,t) ,
\label{axvaten}
\end{eqnarray}

\noindent where $|P, \lambda \rangle$ represents a deuteron state of
momentum $P$ and polarization $\lambda$,  $\bar{P}=(P+P')/2$, and $n^\mu$
is a light-like vector with $n^+=0, \vec{n}_\perp=0$. Due to the spin-one
character of the target, there are more GPDs than in the nucleon case,
but at the same time the set of polarization observables which in
principle could be measured is also richer. 

Our first priority is to obtain some
quantitative estimates of the expected counting rates for some basic
observables like the unpolarized cross section in DVCS, in order to
assess the feasibility of experiments and check which regions are
optimal for measurements.

\begin{figure}[t]
\centerline{\includegraphics[scale=0.4]{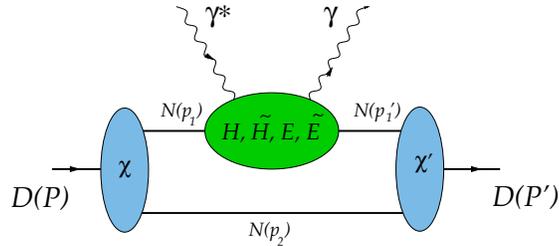}}
\vskip -0.5cm
\caption{Deeply Virtual Compton Scattering off Deuteron in the impulse approximation. The interaction of the two photons with the active nucleon is determined by the nucleon GPDs.}
\end{figure}

\section{THE IMPULSE APPROXIMATION}

Most of the models for the nucleon GPDs take advantage of the known
properties of GPDs in some limits, such as their relationship with
form factors and parton distributions, supplemented with some assumptions
about the $\xi$ dependence. For the deuteron this is a more difficult
task, since nothing is known experimentally about axial form factors
or one of the parton distributions ($b_1$). Moreover, four GPDs
have vanishing first moments and then no information can be inferred
about their $t$ dependence. 

The strategy that we will follow is to use the impulse approximation
to describe DVCS off the deuteron, i.e. to assume that only one nucleon is
active and participates in the absorption and emission of photon, the
other being spectator, Fig. 1. We will also retain only the lowest
Fock-space state and neglect other components in the deuteron wave
function. Within this approximation it is straightforward to link the light-cone wave function of the deuteron to the wave function obtained in non-relativistic quantum mechanics \cite{FRANKFURT79}. 

 The DNN vertex is described by the wave function $\chi_\lambda
 (\alpha, \vec{k}_\perp, \lambda_1,\lambda_2)$ where $\alpha$ is the
 fraction of '+' momentum carried by one of the nucleons and
 $\vec{k}_\perp$ is its transverse momentum in a frame where the total
 transverse momentum of the deuteron vanishes. The indexes $\lambda,
 \lambda_i$ refers to the polarization of the deuteron and nucleons
 respectively. Recall that in Light-front dynamics the two nucleons are on-shell but off (light-cone) energy shell.

	The matrix elements (\ref{vaten},\ref{axvaten}), or equivalently, the deuteron GPDs, can be written, in a somewhat symbolic notation, as a convolution between the deuteron wave functions and the isoscalar flavour combination of the appropriate nucleon GPDs:

\begin{eqnarray}
\label{vatenimpulse}
V_{\lambda'\lambda}(x,\xi,t) & = & \chi_{\lambda'}^*(\alpha',\vec{k}_\perp', \lambda_1', \lambda_2') \otimes H,E(x_N,\xi_N,t) \otimes \chi_{\lambda}(\alpha,\vec{k}_\perp,\lambda_1, \lambda_2)  \\
A_{\lambda'\lambda}(x,\xi,t) & = & \chi_{\lambda'}^*(\alpha',\vec{k}_\perp', \lambda_1', \lambda_2') \otimes \tilde{H},\tilde{E}(x_N,\xi_N,t) \otimes \chi_{\lambda}(\alpha,\vec{k}_\perp,\lambda_1, \lambda_2) 
\label{axvatenimpulse}
\end{eqnarray}

\noindent where $\alpha=\frac{p_1^+}{P^+}$,
$\alpha'=\frac{p_1'^+}{P'^+}$, according to the notation of
Fig. 1. We will not consider here the gluon sector of the GPDs and we
will limit ourselves to the quark content of the nucleon. The arguments of the nucleon GPDs are given by  

\begin{eqnarray}
\xi_N & = & \frac{\xi}{\alpha (1+ \xi)-\xi}  \;\; , \\
x_N & = & \frac{x}{\xi} \;  \xi_N \;\; .
\end{eqnarray}

Notice that the variables $x$ and $\xi$ refer to fractions of plus momentum with respect to $\bar{P}^+$ whereas $x_N$ and $\xi_N$ are defined with respect to $(p_1^+ +p_1'^+)/2$. From the relations above one can see that when $x < \xi $ then $x_N < \xi_N $, i.e. when we enter the kinematic region where we are testing the meson distribution amplitude in the deuteron we are actually looking at this region at the nucleon level. This is just a consequence of retaining only the lowest two-nucleon state in the deuteron. 

The skewness parameter $\xi$ determines the  momentum transfer in the longitudinal direction:

\begin{equation}
\Delta^+ \equiv (P'^+-P^+) = -2 \xi \bar{P}^+  \;\; ,
\end{equation}

\noindent and in the generalized Bjorken limit this is entirely fixed
by the kinematics of the virtual photon ($\xi\approx \xbj/2$). In
the impulse approximation, this momentum transfer has to be provided
by the active nucleon, and after that, 
the final state of this active nucleon still
has to fit into the final deuteron. However, the deuteron is a loosely
bound system, which means that one cannot have a very asymmetrical
sharing of longitudinal momentum between the nucleons and therefore
the formation of the coherent final state will be strongly suppressed
in the impulse approximation for large skewness.

	To be more quantitative let us define the longitudinal momentum distribution of the nucleon in the deuteron as:

\begin{equation}
\label{nalpha}
n_\lambda ( \alpha) = \sum_{\lambda_1, \lambda_2} \int
\frac{d\vec{k}_\perp d\beta}{(16 \pi)^3} \; | \chi_\lambda (\beta,
\vec{k}_\perp,\lambda_1,\lambda_2)|^2 \delta (\alpha-\beta) \;\; ,
\end{equation}
 
\noindent which is normalized according to 

\begin{equation}
\int d\alpha \; n_\lambda(\alpha) =1 \;\; .
\end{equation}

In Fig. 2 we show $n_0(\alpha)$ evaluated with the wave function from the Paris potential \cite{LACOMBE81}. This distribution is strongly peaked at $\alpha=0.5$ and its width is of the order of the ratio of the binding energy divided by the nucleon mass. 

	In the impulse approximation, the active nucleon after the
	interaction with the photon carries a fraction of 
longitudinal momentum which is given by

\begin{equation}
\alpha ' = \alpha - \frac{\xbj}{1-\xbj} (1 - \alpha) \;\; . 
\end{equation}

In Fig. 2 we plot the difference $\alpha - \alpha'$ as a function of
$\alpha$ and for several values of the skewness. We see that for $\xbj
\gtrsim 0.1$ this difference is larger than the width of the momentum
distribution, and therefore, we will inevitably have  
a too fast or too slow nucleon (in
the longitudinal direction). In this case the central region of
momentum, where a maximal contribution is expected, is missed and then
the 
cross sections will decrease very fast with $\xbj$. In other words,
there is an increasing difficulty in forming a coherent final
state as the longitudinal momentum transfer, i.e. $\xbj$ increases. In
that case other coherent mechanisms, which could involve higher
Fock-space components, will presumably become dominant.
 Not much is known about these states, but it should be
emphasized that the suppression of the diagram of Fig. 1 occurs at
$\xbj$ as low as 0.2, so that there is room to check the importance of
the contribution of these 'exotic' states.

	To evaluate the matrix elements (\ref{vaten},\ref{axvaten}) according to the expressions (\ref{vatenimpulse},\ref{axvatenimpulse}) we have used the parameterization of the deuteron wave function provided by the Paris potential \cite{LACOMBE81} which contains a s-wave and a d-wave component. For the nucleon GPDs we have considered only the contribution of $H$ and $\tilde{H}$ since $E$ and $\tilde{E}$ go with suppressing kinematical prefactors. The flavour combination which enters in (\ref{vatenimpulse},\ref{axvatenimpulse}) is the isoscalar one so that there is no dangerous pion pole contribution to $\tilde{E}$. 

To model $H$ and $\tilde{H}$ we have used a phenomenological
parameterization based on double distributions \cite{RADYUSHKIN99},
without considering any D-term at this stage. The $t$ dependence has
been taken into account in a factorized form by multiplying the double
distributions by the corresponding quark form factors extracted from
empirical (dipole) parameterizations of nucleon form factors.  Notice
that even if we take a factorized t-dependence for the nucleon GPDs,
the resulting t-dependence in the deuteron GPDs, obtained through the
expressions (\ref{vatenimpulse},\ref{axvatenimpulse}), cannot be
factorized out and, in fact, the stronger $t$ dependence
comes from the $\vec{k}_\perp$ in the deuteron wave function.
 
	To evaluate the differential cross sections for the process $l
	d \longrightarrow l d \gamma$ we also need to calculate the
	contribution of the concurring Bethe-Heitler (BH)
	mechanism. The amplitude for this process can be calculated
	exactly and the resulting squared amplitude is written in
	terms of the elastic structure functions of the deuteron, which also
	appear in the elastic reaction $e d \longrightarrow e d$ 
\cite{GARCON01}: 

\begin{equation}
|T^{BH}|^2 = 
\frac{(4 \pi \alpha_{\mbox {\tiny em}})^3}{t^2} [ K_a A(t) + K_b B(t)]
\;\; ,
\end{equation}

\noindent where $K_a$ and $K_b$ are kinematical coefficients that will be detailed elsewhere. 

\begin{figure}[t]
\begin{tabular}{lr}
\includegraphics[scale=0.38]{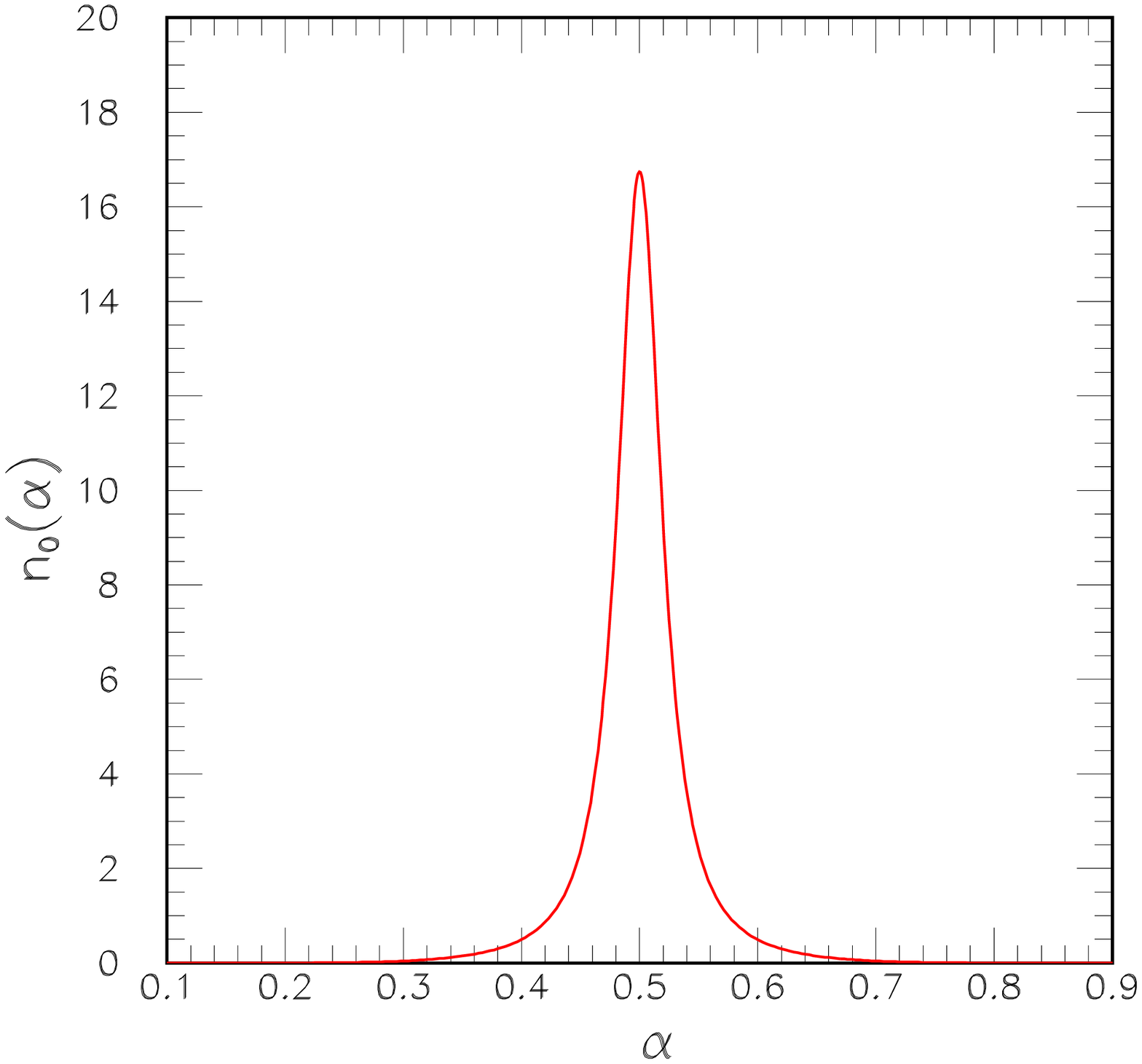} &
\includegraphics[scale=0.38]{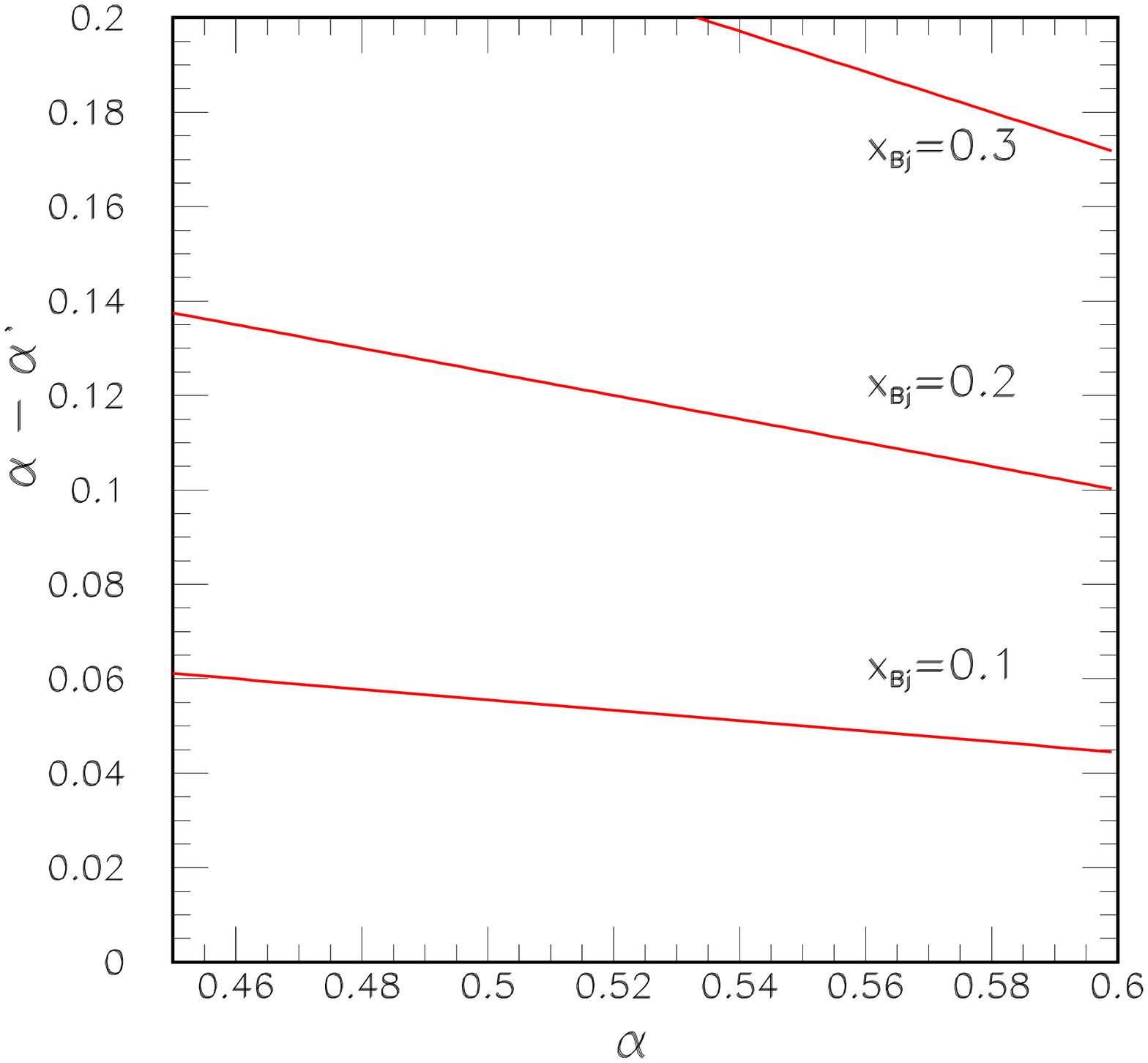} 
\end{tabular}
\vskip -0.5cm
\caption{Longitudinal momentum distribution of the nucleons in the
deuteron with polarization $\lambda=0$ (left). Difference
between final ($\alpha'$) and initial fractions  ($\alpha$) 
of longitudinal momentum carried by the active nucleon 
for different values of $\xbj$ (right).}
\end{figure}

\begin{figure}[t]
\begin{tabular}{lr}
\includegraphics[scale=0.45]{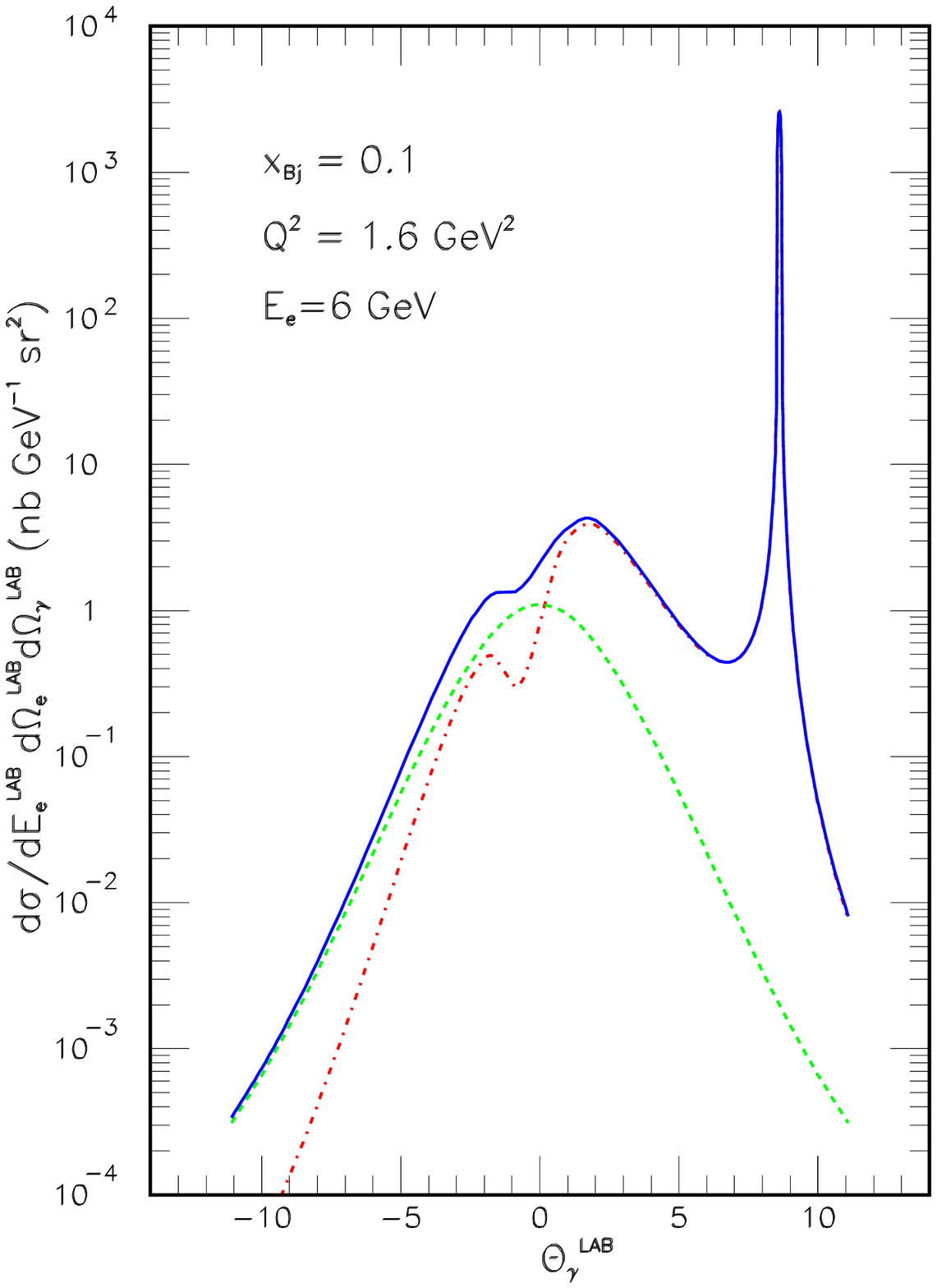} &
\includegraphics[scale=0.45]{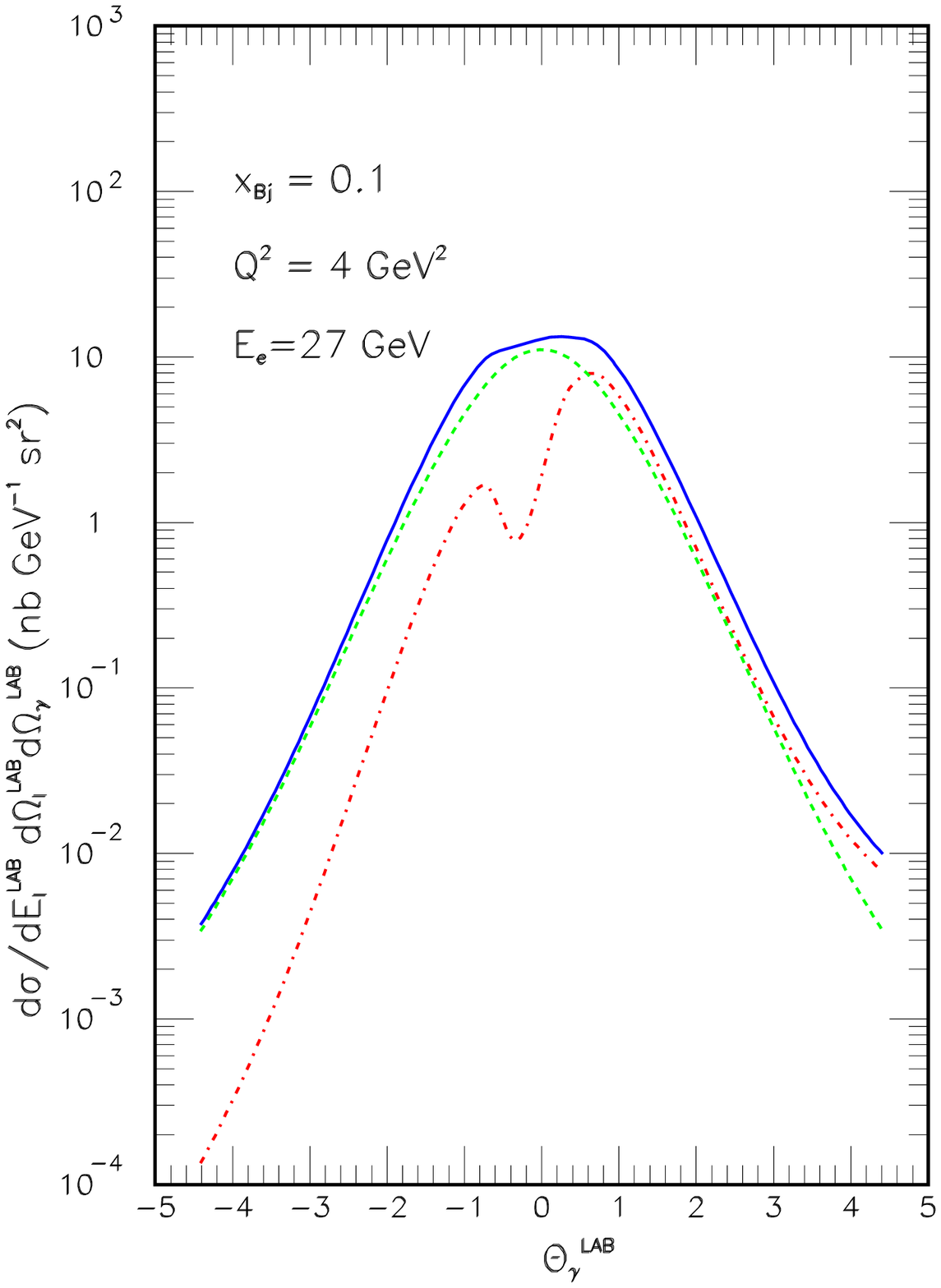} 
\end{tabular}
\vskip -0.5cm
\caption{Unpolarized Differential Cross section for DVCS for typical
kinematics at JLab (left) and HERMES (right).
Dashed-dotted line: BH only; dashed line: VCS only; full
line: BH + VCS + Interference.}
\end{figure}

\section{RESULTS}

In Fig. 3 we show the unpolarized differential cross section for kinematical regions which can be accessed at JLab and HERMES. We see that the cross sections are of the order of the nb. in the central region, with a very fast fall-off. The order of magnitude of the cross sections in this region is comparable to the one obtained for the proton. 

Concerning the separate contributions of BH and VCS we observe also the same patterns as for the nucleon case \cite{VDH99}, with a dominance of the VCS contribution in the region of negative $\theta_\gamma^{\mbox {\scriptsize LAB}}$ (for the in-plane production it corresponds to the case where the scattered lepton and deuteron are detected on the same side of the detector). For the positive $\theta_\gamma^{\mbox {\scriptsize LAB}}$ region, the balance between BH and VCS strongly depends on the energy of the lepton. 

This similarities with the nucleon case leads one to think that we will also have for the deuteron sizeable values for other observables like the beam spin asymmetry, which can be measured more easily than the normalization in any case. Recently, an estimate was made in \cite{KIRCHNER02} based on some simplifying assumptions, leading to a beam spin asymmetry of $-13 \sin \phi$ for the kinematics of HERMES shown in fig. 3, and comparable to the one obtained for the nucleon under the same kinematical conditions. Our estimates within the impulse approximation confirm asymmetries roughly of this size though with an important $\sin 2 \phi$ component.   
  
One important issue in DVCS off deuteron is the dynamical target mass corrections in the VCS sector. At $Q^2=4$ GeV$^2$ these corrections could be moderate but at $Q^2=1.6$ GeV$^2$ they may be important. This problem was addressed in \cite{BELITSKY01} for a spin-1/2 target but nothing is known for a spin-1 target.

	As stated above, at larger $\xbj$ the cross sections die away
	very rapidly though observables like the beam spin asymmetry
	remain still sizeable. For JLab kinematics, 
        it would allow to reach larger
	$Q^2$ of the order or 2-3 GeV$^2$ and stay on a less dangerous
	ground with respect to target mass corrections.

\vskip 0.5cm
We acknowledge useful discussions with M. Diehl, M. Gar\c{c}on and F. Sabati\'e.

\end{document}